%% file: ms.tex
\documentclass{emulateapj}

\slugcomment{Accepted by the Astrophysical Journal}

\shorttitle{Radio Searches of Four AXPs}
\shortauthors{Crawford, Hessels, \& Kaspi}

\newcommand{\transaxp}{XTE~J1810$-$197}

\begin{document}

\title{Deep Searches for Radio Pulsations and Bursts
from Four Southern Anomalous X-ray Pulsars}

\author{Fronefield Crawford\altaffilmark{1}, Jason
W. T. Hessels\altaffilmark{2,3}, \& Victoria M. Kaspi\altaffilmark{2}}

\altaffiltext{1}{Department of Physics and Astronomy, Franklin and
Marshall College, P.O. Box 3003, Lancaster, PA 17604; email:
fcrawfor@fandm.edu}

\altaffiltext{2}{Department of Physics, McGill University, Montreal,
QC H3A 2T8, Canada}

\altaffiltext{3}{Astronomical Institute ``Anton Pannekoek'',
University of Amsterdam, Kruislaan 403, 1098 SJ Amsterdam, The
Netherlands}

\begin{abstract}
We have searched for persistent radio pulsations, bright single
pulses, and bursts from four Southern anomalous X-ray pulsars
(AXPs). Deep observations were conducted at 1.4\,GHz in 1999 July and
August with the Parkes 64-m telescope.  For all of the target AXPs,
the upper limits on integrated pulsed emission are $\sim0.02$\,mJy,
while the limits on the flux of single pulses are $\sim1$\,Jy or
better.  The corresponding radio luminosity limits are significantly
below the observed luminosities of the majority of the observed radio
pulsars, and they are significantly below the radio luminosity of
\transaxp, the only AXP known thus far to emit radio pulsations, at
the epoch when radio pulses were first detected from this source.  Our
null results support the hypothesis that pulsed radio emission from
AXPs is only present in conjunction with X-ray outburst activity.
However, we cannot rule out the possibility that pulsed radio emission
is present, but that it is too weak to be detected in our
observations, or that unfavorable viewing geometries prevent this
emission from being seen by terrestrial observers. Given the possible
association between the radio emission and transient X-ray behavior of
\transaxp, continued radio searches of these and other AXPs at
different epochs are warranted, particularly after periods of X-ray
burst activity.
\end{abstract}


\section{Introduction} 

Anomalous X-ray Pulsars (AXPs), thought to be magnetars
\citep[see][for a review]{wt06,kas06}, are a rare class of slowly
rotating ($P_{\rm spin} = 5-12$\,s) neutron stars whose X-ray emission
may be powered by the decay of their huge inferred magnetic fields
($B_{\rm surf} = 10^{14} - 10^{15}$\,G).  Though they are relatively
bright X-ray sources ($10^{34} - 10^{36}$\,erg~s$^{-1}$), previous
attempts to detect radio pulsations from these objects have been
unsuccessful \citep[e.g.][]{ims02,cpkm02,bri+06}.  Recently, however,
\citet{crh+06} reported the discovery of pulsed radio emission from
the transient AXP \transaxp\ \citep{ims+04} in 2006 March, subsequent
to the discovery of a radio point source in VLA imaging of the area
\citep{hgb+05}.  These discoveries have provided an exciting new
spectral window on magnetar physics and another link between the AXPs
and conventional, rotation-powered radio pulsars \citep[especially the
small population of radio pulsars with high inferred magnetic fields,
e.g.][]{msk+03}. Such a connection was previously suggested by the
similarity seen in the timing behavior of these two classes of
objects, namely the presence of timing noise and glitches
\citep[e.g.][]{kcs99,klc00}.

Interestingly, radio emission from \transaxp, which has shown
tremendous X-ray brightening\footnote{Note that other AXPs have also
shown flux variation \citep[e.g. 1E~2259+586,][]{wkt+04}, and the most
recently established AXP, CXO~J164710.2$-$455216 \citep{mcc+06}, had
an outburst on 2006 September 21 \citep{kbc+06}.}, has only been
observed subsequent to its one known X-ray outburst, which occurred
sometime between 2002 November and 2003 January \citep{ims+04}. This
indicates that its radio emission is not a stable phenomenon, but
rather is somehow associated with its X-ray activity.  Archival radio
pulsation searches prior to this X-ray outburst, in which the AXP was
not detected in radio, show that the pulsed radio flux of \transaxp\
increased by more than an order of magnitude from its quiescent level
\citep{crh+06}.  Current radio monitoring \citep{ccr+07} reveals that
there remains an intrinsic flux variability of about a factor of two
on day timescales, and that there has been a gradual decrease in
brightness and a broadening of the pulse profile since its discovery
as a pulsed radio source.  Such large-scale intrinsic flux variability
has come as a surprise and is not seen in normal radio pulsars
\citep{ssh+00}.  \transaxp\ is also peculiar in that it has a very
flat radio spectrum and was at the time of its discovery the brightest
radio pulsar in the sky above 20\,GHz.

If the recent emergence of radio emission from \transaxp\ is indeed
directly related to its X-ray behavior, then that may explain why the
``persistent'' AXPs\footnote{See the McGill University magnetar
catalog at
http://www.physics.mcgill.ca/$\sim$pulsar/magnetar/main.html.} are (at
the very least) radio dim.  On the other hand, the small beaming
fraction of long-period radio pulsars, presumably owing to their large
light cylinders and correspondingly small polar caps
\citep[e.g.][]{ran90,ymj99}, and the narrow radio beam of \transaxp\
itself \citep{crh+06}, suggest that AXPs, if they are generically radio
emitters, should have narrow emission beams and therefore would be
unlikely to be detected.  In the latter case, that the only confirmed
transient AXP is also the only known radio-emitting AXP could be a
geometrical coincidence, although it would still be unclear why it was
not seen before its X-ray outburst.  Only additional radio
observations of the AXPs can distinguish between these possibilities
and determine the phenomenological link between X-ray and radio
activity.  Furthermore, given the observed radio flux variability of
\transaxp\ on both day and year timescales, {\it repeated}
observations of the AXPs are certainly warranted.

Here we report a sensitive search for radio emission, both periodic
and bursting, from three southern AXPs and one southern AXP candidate
(potentially also a transient AXP) using the Parkes radio telescope.
We detect no emission of either kind and provide the most stringent
upper limits to date on such emission from these objects.  Our
results, together with those recently reported by \citet{bri+06} for
these same objects, establish that any periodic or bursting radio
emission from these sources is either non-existent, very weak, or very
sporadic at the epoch of these observations, as compared with
\transaxp, most of the known rotating radio transients
\citep[RRATs,][]{mll+06}, and the bulk of the known radio pulsar
population.  These null results lay the foundation for future radio
observations of these AXPs in the event that one exhibits a
significant X-ray outburst like that seen in \transaxp, as this could
in principle establish whether AXP radio emission is associated in
some way with X-ray outbursts.

\section{Observations} 

Three established AXPs (1E 1048.1$-$5937, 1E 1841$-$045, and 1RXS
J170849.0$-$400910) and one AXP candidate (AX J1845$-$0258\footnote{AX
J1845$-$0258 is a transient AXP candidate only, see \S4 for further
discussion.}) were observed with the Parkes telescope in 1999 July and
August (Table~\ref{tbl-1}) as part of a larger search campaign to find
young, energetic radio pulsars \citep{cpkm02}.  To our knowledge, none
of these sources was in a state of X-ray outburst at the time of our
observations, although such outbursts are known to be short-lived and
some may have been missed, despite the X-ray monitoring of these
objects \citep[e.g.][]{kcs99,kgc+01}.  
Each AXP was observed continuously with a single
16800-s (4.7-h) integration using the center beam of the multibeam
receiver \citep{swb+96, mlc+01}. This receiver has been successfully
used in a number of targeted pulsar searches
\citep[e.g.][]{cmg+02,crg+06} and large-scale pulsar surveys
\citep{mlc+01,ebvb01,bjd+06}.  The central observing frequency was
1374\,MHz with 288\,MHz of bandwidth, consisting of 96 3-MHz
channels. Each channel was 1-bit sampled every 0.25\,ms and recorded
onto DLT magnetic tape.  At the start of each observing session, a
bright test pulsar (PSR J1359$-$6038) was observed to verify that the
observing system was working properly.  Details of the observing setup
are similar to those of the Parkes Multibeam Survey \citep{mlc+01} and
the radio search of these same four AXPs by \citet{bri+06}. Although
the observations of \citet{bri+06} were conducted with the same
observing setup as ours, our observations were longer, analyzed
somewhat differently, and, most importantly, were taken at a different
epoch.  The latter point is particularly relevant given the transient
nature and day-to-day variability seen in \transaxp.

\section{Data Analysis and Results}

Each AXP observation was searched for periodicities, single pulses,
and bursts, and was folded at the known neutron star spin period using
the X-ray timing ephemeris when available.  Standard Fourier-based
periodicity searches were reported by \citet{cpkm02}; no AXP
counterparts or periodicities indicative of another previously
undiscovered pulsar in the same field were found in those searches.
Here we describe the single pulse/burst and direct folding analyses,
which were conducted with the {\tt PRESTO}\footnote{Available at
http://www.cv.nrao.edu/$\sim$sransom/presto.} suite of pulsar analysis
tools \citep{ran01, rem02}.

The first step in both of these analyses was to excise narrow-band and
transient radio frequency interference (RFI) from the data by
identifying bad frequency channels or time intervals and then masking
these. Given the slow rotation rates of the AXPs, and the pernicious
effect of low frequency, ``red'' noise in radio data, this step was
particularly important for ensuring that a relatively weak signal was
not being hidden by RFI.  Typically 15-25\% of the frequency channels
and $\sim 5$\% of the integration time were masked in this process.

\subsection{Folding Search}

In the folding search, the entire 288\,MHz of observing bandwidth was
subdivided into 32 9-MHz subbands, each of which was transformed to
the Solar System barycenter and dedispersed at a dispersion measure
(DM) of 500\,pc cm$^{-3}$.  Each subband was folded at the precisely
predicted spin period of the AXP at the observation epoch using a
contemporaneous ephemeris from X-ray observations (see
Table~\ref{tbl-1} for references; no ephemeris was available for
AX~J1845$-$0258 so the discovery period was used).  Trial DMs were
created by shifting the subbands with respect to each other in pulse
phase. DMs were searched in spacings of $36-72$\,pc cm$^{-3}$ from zero  
up to a maximum DM of $4000-8000$\,pc cm$^{-3}$, depending on the spin
period.  We also folded the data allowing for a search in period which
ranged $\pm 5-10$\,ms from the predicted pulse period.  In the case of
AX J1845$-$0258, we searched a period range between its discovery
period and a period 212 ms greater than this. The upper limit of this
range was determined by using twice the period derivative that results
from assuming that AX J1845$-$0258 has a magnetic field equal to that
of SGR J1806$-$20, the magnetar with the largest inferred field.
Separate folds were also made for each target from nine overlapping
shorter segments of the full observation. These segments started at
intervals of 10\% of the data length and consisted of 20\% of the
total integration time. This was done to account for the presence of
significant scintillation (unlikely given the high predicted DMs of
these sources), strong transient RFI near the AXP spin period, or
large-scale intrinsic pulse strength variability on a timescale
shorter than the observation length.

We found no convincing periodic signals from any of the AXPs in any of
the folds we made.  Apart from the statistical significance of the
measured $\chi^{2}$ values in the folded profiles, the criteria for
differentiating signals potentially related to the AXPs from RFI were
that the AXP signal should be broadband, have a steady pulse phase,
and be dispersed.  None of the observed signals met these criteria,
and we therefore associate them with RFI.  We note that the frequency
decimation of the data into subbands, which required an assumed DM,
did not introduce a significant amount of dispersive smearing compared
to the long spin periods of these neutron stars. Furthermore,
dispersive smearing within the frequency channels and multipath
scattering effects are not likely to be significant factors in our
non-detections: the estimated smearing and scattering times (obtained
from the NE2001 Galactic electron model of \citet{cl02} assuming DM
values up to $\sim 2000$ pc cm$^{-3}$) are less than 1.3\% of the
pulse period in each case. This is significantly less than the 2.7\%
duty cycle assumed in our sensitivity calculations, which is the duty
cycle of \transaxp's integrated radio profile in its discovery paper
\citep[see Figure 1 of][]{crh+06}.

We used the modified radiometer equation, presented by \citet{dtws85}
and more recently by \citet{lfl+06}, to estimate the 1400-MHz upper
limits on the pulsed radio emission from these AXPs.  This equation is
appropriate to use for the folding searches we conducted \citep[see,
e.g., Figure 2 of][which shows the fidelity of using this equation for
folding-search sensitivity estimates]{lfl+06}.  The sky temperature at
1374\,MHz was estimated for each target AXP using the model of
\citet{hssw82} and assuming a radio spectral index of $-2.6$.  As
stated above, a duty cycle of 2.7\% was also assumed. The sensitivity
limit can easily be computed for different assumed duty cycles in the
modified radiometer equation. High-pass filtering is also present in
the observing system; this filter has a time constant of $\sim 0.9$\,s
\citep{mlc+01}. However, if the radio pulses from our target sources
are as narrow as those seen from \transaxp\ ($\la 0.3$\,s for all four
of our targets, assuming a 2.7\% duty cycle), many harmonics of the
fundamental will not suffer any filtering.  We therefore do not
include any attenuation factor from this filtering in our sensitivity
estimates, in contrast to \citet{bri+06}. If our assumption is
correct, then \citet{bri+06} have underestimated the true sensitivity
of their observations by a factor of $2-2.5$.  Table \ref{tbl-1} lists
our calculated upper limits for pulsed radio emission at 1400\,MHz. It
should be noted that these limits do not include loss of data from RFI
masking effects and do not account for any red noise contribution from
RFI, which is time-variable and thus difficult to quantify precisely.
Roughly speaking, however, we estimate that the RFI environment may
worsen these limits by a factor of two.  We also used the estimated
distances to the AXPs to calculate the corresponding 1400-MHz
luminosity limits (Table \ref{tbl-1}).

\subsection{Single Pulse Search}

To search for dispersed single pulses and bursts, the raw data were
dedispersed at a set of trial DMs ranging from zero to $4000-8000$\,pc
cm$^{-3}$, depending on the spin period.  The DM spacing varied as a
function of DM (being wider at higher DMs) and was chosen so as not to
increase the dispersive smearing already present in the frequency
channels. Using the native 0.25-ms time resolution of the data, each
dedispersed time series was searched for candidate single pulses
having a signal-to-noise ratio (S/N) greater than 6.5.  This S/N
threshold was chosen to avoid confusion with the RFI background. To
maintain sensitivity to pulses broader than 0.25\,ms,
matched-filtering with a boxcar function of varying widths (ranging
from 1 to 30 samples) was also done.\footnote{The boxcar widths used
were 1, 2, 3, 4, 6, 9, 14, 20, and 30 samples.} To enhance sensitivity
to even longer-duration single pulses and bursts, the dedispersed time
series were then downsampled by combining the original samples into
contiguous blocks of 2, 4, 8, 16, and 32 samples. Each set of
downsampled time series was then searched using the same set of boxcar
filters, providing sensitivity to signals of duration $\la 240$\,ms
(this is equal to the maximum boxcar length, 30 bins, times the
maximum downsampled sample time, 8\,ms). This range encompasses and
significantly exceeds the $\la 10$\,ms single-pulse widths seen by
\cite{crh+06} for \transaxp. The sensitivity to pulses in the
downsampled data increases as the square root of the number of
combined samples (and further increases with the boxcar filtering),
but for each downsampling, a slightly higher S/N threshold was used to
eliminate spurious candidates from RFI. Specifically, we used S/N
thresholds of 7.0, 8.0, 9.0, 11.0, and 12.0 for the downsampling
factors described above. To differentiate dispersed single pulses from
RFI, results from the different trial DMs were stacked on top of each
other in order to identify signals that peaked in S/N at a non-zero DM
\citep[e.g.,][]{lk04,mll+06}.  No candidate signals were detected from
any of the target AXPs in any of these search trials. Throughout the
search, we maintained excellent sensitivity to single pulses for all
reasonable DMs.

We have determined single-pulse flux density and luminosity detection
limits over the range of pulse time scales we considered ($0.25 -
240$\,ms), using the modified radiometer equation (e.g., Lorimer et
al.  2006\nocite{lfl+06}) and the S/N thresholds described above.
These limits are presented in Table \ref{tbl-1}.  As in the case of
the folding search, the deleterious effects of red noise and data loss
from RFI masking were not included in the sensitivity limits.

\section{Discussion and Conclusions} 

We place a flux density limit of $\sim 0.02$\,mJy on integrated pulsed
radio emission from our four AXP targets.  This is about ten times
lower than the upper limits placed on \transaxp's pulsed emission in
archival Parkes Multibeam Survey observations prior to the outburst of
that source \citep[see Table~1 in][]{crh+06}. Our greater sensitivity
compared with the Parkes Multibeam Survey data is the result of $8
\times$ longer integration time and higher gain because our sources
were specifically placed at the center of the receiver beam. However,
the corresponding 1400-MHz luminosity limits from our observations are
only $2-3 \times$ lower than those established for \transaxp\ prior to
outburst because that source is estimated to be $2-3 \times$ closer
than our target AXPs. The public pulsar catalog
\citep{mhth05}\footnote{Available at
http://www.atnf.csiro.au/research/pulsar/psrcat.} shows that of the
known pulsars with measured 1400-MHz radio luminosities, $\sim 5$\%
have values below 1\,mJy kpc$^{2}$, which is comparable to our
limits. Based on the work of \citet{lml+98}, \citet{gsgv01} estimate
that $\ga 60$\% of potentially observable pulsars have 1400-MHz
luminosities less than 1\,mJy kpc$^{2}$.  It is therefore conceivable
that weak radio pulses could be emitted by our targets but that they
are below our detection threshold.

However, the average 1400-MHz flux density reported by \citet{crh+06}
for \transaxp\ soon after it was first detected was $\sim
8$\,mJy. This corresponds to a 1400-MHz luminosity of $\sim 80$\,mJy
kpc$^{2}$, which is almost two orders of magnitude larger than the
luminosity limits of our searches. Thus, comparably strong radio
emission from our target AXPs would have been easily detectable if it
were beamed in our direction.  Long-period pulsars typically have
smaller duty cycles and narrower emission beams than faster pulsars
\citep{ymj99}, thereby making line-of-sight detection more unlikely.
It is possible that all of our target AXPs could be strong radio
emitters but that none is detectable by us owing to unfavorable
viewing geometries.  We note however, that the emission from
\transaxp\ is different than for normal rotation-powered radio
pulsars. \citet{crh+06} argue that it is possible that the radio
emission from \transaxp\ is generated on closed field lines (as
opposed to conventional radio pulsar emission, which streams along
open magnetic field lines above the magnetic poles) and beamed into a
much broader range of angles.  If radio emission from magnetars is
emitted in a variety of directions, then detectability is less
contingent on beaming.

One of the sources in our sample, AX~J1845$-$0258, is a candidate
transient AXP having had an X-ray outburst in 1993 \citep{gv98,tkk+98}
but not having been seen to emit pulsed X-ray emission since then. In
this sense, it may be like \transaxp, but in a quiescent state in
1999, when the observations reported here were taken \cite[see][for a
more detailed summary of this source's X-ray behavior]{tkgg06}.  If
AX~J1845$-$0258 is indeed an AXP and undergoes another flare in the
future, it will be important to compare any subsequently detected
radio emission with that seen from \transaxp\ subsequent to its X-ray
outburst.  In that event, the upper limits presented here during
AX~J1845$-$0258's quiescent phase will be a critical piece of
information.  The observation of an outburst on 2006 September 21
\citep{kbc+06} from the most recently established AXP,
CXO~J164710.2$-$455216 \citep{mcc+06,wkg06}, suggests that this
outburst was similar to the one seen from \transaxp.  No pulsed radio
emission was detected a week after the burst, to a limit of 0.04\,mJy
\citep[][further observations are planned]{brip06}.

For a comparison of our single-pulse upper limits with the observed
single pulses reported for \transaxp\ by \citet{crh+06}, we compared
the corresponding single-pulse luminosity limits for our sources and
for \transaxp\ (Table~\ref{tbl-1}). Our luminosity limits in the most
conservative case range from 22 to 69 Jy kpc$^{2}$, below the
luminosity of $\ga 100$\, Jy kpc$^{2}$ derived from the $\ga 10$\,Jy
pulses for \transaxp\ reported by \citet{crh+06}. Single pulses were
also detected from almost every rotation of \transaxp.  This suggests
that we would have expected to detect a large number of pulses during
the course of our observations if comparable radio emission were being
emitted by our targets.

Our non-detection of single pulses further weakens the hypothesis that
some of the sporadically emitting radio sources known as RRATs
\citep{mll+06} are magnetars.  This link has also been weakened by two
other recent results.  First, the discovery of X-ray emission
associated with the most ``magnetar-like'' of the RRATs, J1819$-$1458,
shows that this emission is more typical of middle-aged
rotation-powered pulsars than that seen from the magnetars
\citep{rbg+06}.  Secondly, \citet{wsrw06} argue that the nearby
rotation-powered pulsar PSR~B0656+14 shows emission that would cause
it to be identified as an RRAT if its distance were greater. We note,
however, that single radio pulse/burst searches towards the AXPs are
also motivated for other reasons.  \transaxp\ has shown strong
individual pulses, with an intensity distribution that follows a power
law.  This presents the possibility of detecting individual bright
pulses even if the majority of the radio emission is below the
detection threshold.  If magnetars can emit giant pulses, this is also
potentially interesting, as such pulses may be generated in a
different part of the magnetosphere than ``normal'' pulses
\citep[][and references therein]{kbm+06}.  If beaming plays an
important role in the detectability of radio emission from magnetars,
then giant pulses could provide another window in which to detect
radio emission.  Furthermore, AXPs are known to emit rare X-ray
bursts.  One of the AXPs in our sample, 1E~1048.1$-$5937, was the
first shown to have such behavior \citep{gkw02}.  Such X-ray bursts
may be accompanied by radio bursts \citep{l02}.

In conclusion, our non-detections support the hypothesis that pulsed
radio emission from AXPs is connected to X-ray outburst activity, as
may be the case for \transaxp.  However, unfavorable viewing
geometries and very weak radio emission cannot be dismissed as reasons
for our non-detections.  Given that the transient radio emission
observed from \transaxp\ appears to be linked with its transient X-ray
behavior, continued radio monitoring of all the AXPs is warranted,
particularly after periods of X-ray burst activity.

\acknowledgements

We thank Scott Ransom for creating the {\tt PRESTO} suite of software
tools, which were used in this analysis, and we acknowledge use of the
McGill University magnetar catalog.  J.W.T.H. holds an NSERC
Postdoctoral Fellowship.  V.M.K. is the Lorne Trottier Canada Research
Chair and acknowledges support from NSERC, CIAR, and FQRNT.  The
Parkes radio telescope is part of the Australia Telescope, which is
funded by the Commonwealth of Australia for operation as a National
Facility managed by CSIRO.

\bibliographystyle{apj}
\bibliography{journals1,modrefs,psrrefs,chk07_ref}

\clearpage

\input{tab1}

\end{document}

%% file: tab1.tex
\tabletypesize{\scriptsize}
\begin{deluxetable}{lcccc}
\tablecaption{Radio Search Parameters and Results.\label{tbl-1}}
\tablewidth{0pt}
\tablehead{
\colhead{} &
\colhead{1E 1048.1$-$5937} &
\colhead{AX J1845$-$0258} &
\colhead{1E 1841$-$045} &
\colhead{1RXS J170849.0$-$400910}
}
\startdata
Spin period (s)       &  6.45    & 6.97    & 11.77   & 11.00  \\
Ephemeris Reference       & \citet{kgc+01} & \citet{tkk+98}\tablenotemark{a} & \citet{gvd99} & \citet{gk02} \\
Galactic longitude, latitude (deg.) & 288.26, $-$0.52 & 29.52, 0.07  & 27.39, $-$0.01 & 346.47, 0.03 \\
$T_{\rm sky}$ (K)\tablenotemark{b}    & 9.1      & 12.3    & 13.2    & 16.3   \\
Observation MJD              & 51378    & 51391   & 51382   & 51379  \\ 
Observation date             & 1999 Jul 19 & 1999 Aug 1 & 1999 Jul 23 & 1999 Jul 20 \\
$S_{1400}$ (mJy)\tablenotemark{c}     & $\la 0.02$  & $\la 0.02$ & $\la 0.02$  & $\la 0.02$  \\
$S_{1400}$ single (mJy)\tablenotemark{d} & $\la 875 - 50$ & $\la 975 - 60$ & $\la 1000 - 60$   & $\la 1085 - 65$   \\

Distance (kpc)\tablenotemark{e}       & $\sim 5$?  & $\sim 8$?  & $\sim 7$ & $\sim 8$?  \\
$L_{1400}$ (mJy kpc$^{2}$)\tablenotemark{f} & $\la 0.5$   & $\la 1.3$    & $\la 1.0$     & $\la 1.3$   \\
$L_{1400}$ single (Jy kpc$^{2}$)\tablenotemark{g} & $\la 22 - 1.3$ & $\la 62 - 3.7$ & $\la 49 - 2.9$ & $\la 69 - 4.1$ \\
\enddata

\tablecomments{Each observation was 16800\,s long with 0.25-ms sampling and 
288\,MHz of bandwidth centered at 1374\,MHz. These values were used in
the sensitivity estimates presented below. Including RFI excision effects
in these estimates increases the limits by $\sim 10$\%.}

\tablenotetext{a}{There is no period derivative available for this candidate
AXP.}

\tablenotetext{b}{1374-MHz sky temperature estimated from the model of
\citet{hssw82}, assuming a spectral index of $-2.6$.} 

\tablenotetext{c}{1400-MHz flux density limit 
on pulsed emission estimated from the modified radiometer equation
\citep{dtws85, lfl+06}. A pulsed duty cycle of 2.7\%  and
a S/N threshold of 7 was assumed in each case. No red noise
contribution due to RFI was included in these estimates.}

\tablenotetext{d}{Range of single-pulse 1400-MHz 
flux limits for the range of time scales searched ($0.25 -
240$\,ms). No red noise contribution from RFI was included in these
estimates.}

\tablenotetext{e}{Distances were taken from Table~1 of \citet{bri+06}, 
which were based on associations described by \citet{gsgv01} and
\citet{gmo+05}. Question marks indicate a significant uncertainty in the
actual value owing to different distance estimates by different
authors (e.g., Durant \& van Kerkwijk 2006\nocite{dv06a}; Kuiper et
al. 2006\nocite{khd+06}).}

\tablenotetext{f}{1400-MHz luminosity limit on pulsed emission. This 
assumes a beaming fraction of one sr, so that $L_{1400} = S_{1400}
d^{2}$, where $d$ is the best estimate of the AXP distance.}

\tablenotetext{g}{1400-MHz luminosity limits on single pulses as
determined from the 1400-MHz single-pulse flux limits and the best
estimate of the AXP distance (see also the above footnote).}

\end{deluxetable}